\documentclass[final,5p,times,twocolumn]{elsarticle}

\usepackage{mypreamble}

\usepackage{silence}
\begin{document}

\begin{frontmatter}

\title{An Effective Technique for Increasing Capacity and Improving Bandwidth in 5G NB-IoT}

\tnotetext[mytitlenote]{Corresponding author.}
\author[1]{Abdulwahid Mohammed \corref{mycorrespondingauthor}}
\ead {abdulwahid.21@azhar.edu.eg}

\author[2,3]{Hassan Mostafa}
\ead{hmostafa@zewailcity.edu.eg}

\author[1]{Abd El-Hady. A. Ammar}
\ead{hady42amar@gmail.com}

\address[1]{Electronics and Communication Engineering Department, Al-Azhar University, Cairo 11651, Egypt.}
\address[2]{Electronics and Communication Engineering Department, Cairo University, Giza 12613, Egypt.}
\address[3]{University of Science and technology, Zewail City of Science and Technology, Giza 12578, Egypt.}

\begin{abstract}
The Internet of Things (IoT) has changed dramatically in recent years. The number of IoT devices is rapidly expanding, and various new IoT applications relating to automobiles, transportation, power grid, agriculture, metering, and other areas have emerged. With hundreds of billions of connected devices, it is important for researchers to create effective resource management approach to satisfy the quality of service (QoS) requirements of 5th generation (5G) and beyond. Furthermore, wireless spectrum is increasingly scarce as demand for wireless services develops, demanding imaginative approaches to increase capacity within a limited spectral resource in order to meet service demands. In this article, a modified symbol time compression (M-STC) technique is suggested to paves the way for 5G networks and beyond to enhance the capacity and throughput. The M-STC method is a compressed signal waveform technique that increases the capacity by compressing the occupied bandwidth without increasing the complexity, losing data throughput or bit error rate (BER) performance. A comparative analysis is provided between the traditional orthogonal frequency division multiplexing (OFDM) system, OFDM using conventional symbol time compression (C-STC-OFDM) and OFDM using the proposed technique (M-STC-OFDM). 
The simulation results using Matlab-2021a show that the suggested method, M-STC-OFDM, drastically lowers the time needed for each OFDM signal by 75\%. As a consequence, the M-STC-OFDM system decreases bandwidth (BW) by 75\% when compared to a standard OFDM system (BW$\mathrm{_{OFDM}\sim}$ 180 kHz and BW$\mathrm{_{M-STC-OFDM}\sim }$ 45 kHz), while the C-STC-OFDM system reduces BW by 50\% (BW$\mathrm{_{C-STC-OFDM}\sim }$ 90 kHz). Furthermore, using the M-STC-OFDM system reduces peak to- average-power-ratio (PAPR) by 2.09 dB when compared to the standard OFDM system and 1.18 dB when compared to C-STC-OFDM with no BER deterioration. Moreover, as compared to the 16QAM-OFDM system, the proposed M-STC-OFDM system reduces the signal-to-noise-ratio (SNR) by 3.8 dB to transmit the same amount of data.
\end{abstract}


\end{frontmatter}

    \section{Introduction}
\begin{figure}[!ht]
	\centering
	\includegraphics[width=0.9\linewidth,height= 12. cm]{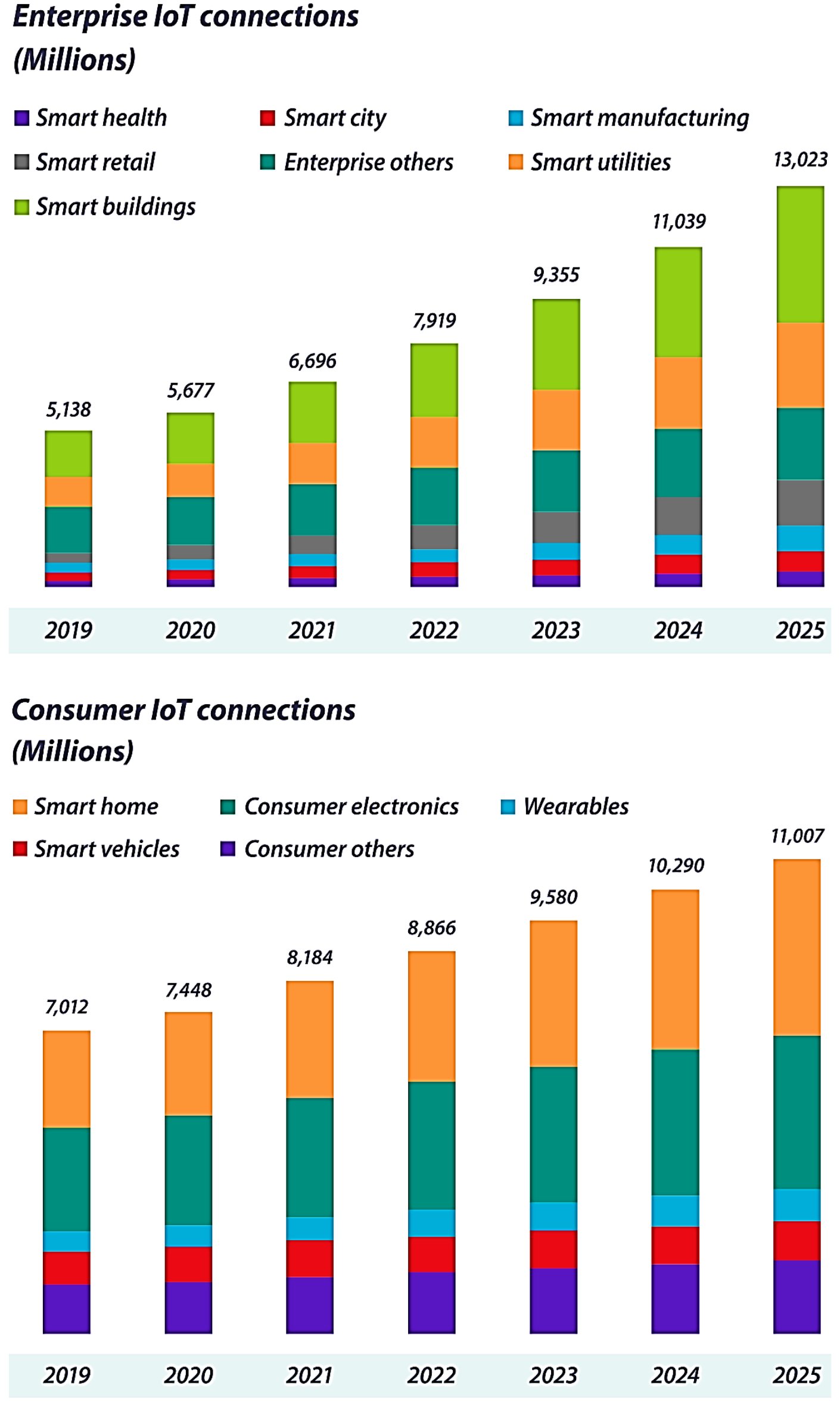}
	\caption{\label{IOT_Consumer} IoT connections (include impact of COVID-19) \cite{Ioconnection}}.	
\end{figure}

In recent years, the Internet of Things (IoT) has evolved tremendously and the number of IoT devices is fast growing. The IoT provides a wide range of possibilities for novel applications aimed to enhance our life quality. The number of IoT devices is rapidly expanding, and various novel IoT applications relating to automobiles, transportation, power grid, agriculture, metering, and other areas have emerged \cite{liu2020analysis}. In particular, the IoT enables humans to live in a smarter environment than ever before. For instance, residents all around the world receive personalized urban services on a continuous, automated, and collaborative basis \cite{zhang2020enabling}. The IoT includes smart devices (such as mobile devices and wearables), next-generation cellular networks (NGCNs), computation infrastructure (such as edge computing and cloud computing), and other core technologies (such as sensing and identification) \cite{arshad2017green}. Moreover, IoT devices have emerged in different environments, such as smart grids \cite{parsa2017implementation}, industrial automation \cite{desima2017alarm},  smart cities, healthcare and home appliances \cite{yang2015intelligent}. According to GSMA Intelligence IoT Connections Forecast (5) and Ericsson Mobility's Report (4), both of which were issued in June 2020 (these two reports includes the impact of the COVID-19 epidemic on the IoT industry), the total number of IoT device connections is expected to approach 24 billion among all IoT markets by 2025 \cite{Ioconnection}. The smart home, consumer electronics, wearables, and smart vehicles sectors, are expected to be the primary drivers of development in consumer IoT, according to the "GSMA Intelligence Report$^6$" \cite{Ioconnection}. Fig. \ref{IOT_Consumer} depicts all sorts of IoT devices from a variety of industries, as well as consumer and business applications. 

A variety of Low Power Wide Area (LPWA) systems have evolved to handle this massive data need. These technologies enable wide coverage, low power consumption, a large number of users, and low device complexity \cite{sari2020load,su2020energy}. Various standardization committees, like as IEEE, 3GPP, and others, are working to standardize LPWA technology. LPWA can employ cellular or non-cellular wireless technologies. Cellular technologies include Machine Type Communication (MTC), enhanced Machine Type Communication (eMTC) and Narrow-band Internet of Things (NB-IoT), whereas non-cellular technologies include Long Range (LoRa), ZigBee, Bluetooth, Z-Wave, and others \cite{rastogi2020narrowband,kisseleff2020efficient,migabo2020narrowband}. With the explosive growth of 5G new radio technologies, industry and academia are focusing their efforts on enhanced mobile broadband (eMBB), massive machine-type communications (mMTCs), and ultra-reliable low latency communications (URLLCs) \cite{liu2019eliminating}. To meet the 5G vision, it is required to not only make substantial advancements in new wireless technologies, but also to consider the harmonic and equitable coexistence of diverse networks, as well as the compatibility of 4G and 5G systems \cite{niu2015exploiting}. \\

The rest of this paper is organized as follows. {\color{teal}Section 2} summarizes the related work and contribution.  {\color{teal}Section 3} describes the overall mathematical model of the proposed technique.  {\color{teal}Section 4} explains the mathematical notion of the PAPR. The simulation results and comments are presented in  {\color{teal}Section 5}.  {\color{teal}Section 6} demonstrates the computational complexity of the different systems. This study is summarized in  {\color{teal}Section 7} by outlining the  advantages of the proposed technique. 
    \section{Related Work and Contribution}
 In \cite{xu2018uplink}, the authors have proposed an NB-IoT architecture based on an sophisticated signal waveform known as non-orthogonal spectrum efficient frequency division multiplexing (SEFDM). Comparing with OFDM, the developed waveform might enhance data rate without requiring extra bandwidth. According to the simulation results, the suggested waveform might enhance data rates by 25\% when compared to the OFDM signal waveform. However,  the non-orthogonality of the sub-carriers may leads to inter carrier interference (ICI), requiring additional power consumption on the receiver side. Since the signal processing is done at the base stations, it is appropriate for the up-link channel \cite{xu2018uplink}. But, this is not practicable since it requires additional processing at the down-link channels.
 
 In \cite{xu2018non,DoublingIoT}, the authors have described solutions  for NB-IoT employing fast-orthogonal frequency division multiplexing (Fast-OFDM) and symbol time compression (STC) based approach, as the two techniques  show their advantages compared to the standard orthogonal frequency division multiplexing (OFDM).  When compared to a typical OFDM system, the Fast-OFDM technique reduces the space between sub - carriers by 50\% and avoids BER degradation.   However, F-OFDM may causes carrier frequency offset (CFO).  Additionally, the PAPR problem still affects this technique.  In addition to the same benefits of the F-OFDM technique, the STC based approach in \cite{DoublingIoT} has the ability to reduce the PAPR and CFO issue.
 
The authors of \cite{xu2019waveform} employ non-orthogonal multi-carrier SEFDM wave-forms for single- and multiple-antenna systems and show how these wave-forms may optimize down-link (DL) bandwidth by 11\% when compared to NB-IoT. The results demonstrate that enhanced NB-IoT (eNB-IoT) has the same efficiency as NB-IoT in both single and multiple antenna for modulation schemes such as 4QAM and 8QAM. However, NB-IoT outperforms  eNB-IoT in higher order modulation formats such as 16QAM.

In this study, M-STC is suggested as a promising and effective technique for 5G systems and beyond for the following reasons. 
\begin{itemize}
	\item Enhances data rate while keeping the  modulation types.
	
	\item Lowers power consumption by reducing the number of sub-carriers needed. 
	
	\item As opposed to Fast-OFDM in \cite{xu2018non}, it does not cause a mismatch  in sampling rate or CFO since the space between the sub-carriers is not reduced.

	\item Enhances system performance by further improving the PAPR issue and keeps the system from degrading in BER.

	\item Unlike the previous work, F-OFDM in \cite{xu2018non} and C-STC-OFDM in \cite{DoublingIoT}, the suggested approach (M-STC-OFDM) is appropriate for huge connections since it decreases the used bandwidth to a quarter (save 75\% of BW) compared to typical OFDM systems, allowing the unused bandwidth to be utilized to transmit additional data. Furthermore, the M-STC-OFDM could transfer twice as much data as F-OFDM and C-STC-OFDM systems.
    
    \item Although employing the M-STC technique  with OFDM systems could compress the bandwidth to a quarter as compared to a conventional OFDM system, it still has the same complexity.
\end{itemize}

\section{System Model}
The C-STC scheme is initially presented in \cite{el2017time}, where it was found to compress the symbol time to half and save 50\% of bandwidth. The C-STC approach is used with OFDM system (C-STC-OFDM) to transmit the same amount of data as the typical OFDM system while utilizing just half the bandwidth. In this article, the M-STC technique is proposed which saves 75\% of symbol time and thereby decreases bandwidth to a quarter.  However, using M-STC does not cause the deterioration in BER or an increase in complexity, as it will be demonstrated in further details in section V and section VI. The M-STC method is implemented on the transmitter side, while the modified symbol time extension (M-STE) technique is implemented on the receiver side. The mathematical model can be divided into three parts as follows: 1) the M-STC mathematical model at the transmitter side, 2) the M-STE mathematical model at the receiver side and 3) the OFDM system using M-STC (M-STC-OFDM).
\subsection{M-STC System Model}
\begin{figure}[!ht]
	\centering
	\includegraphics[width=\linewidth,height= 6.5 cm]{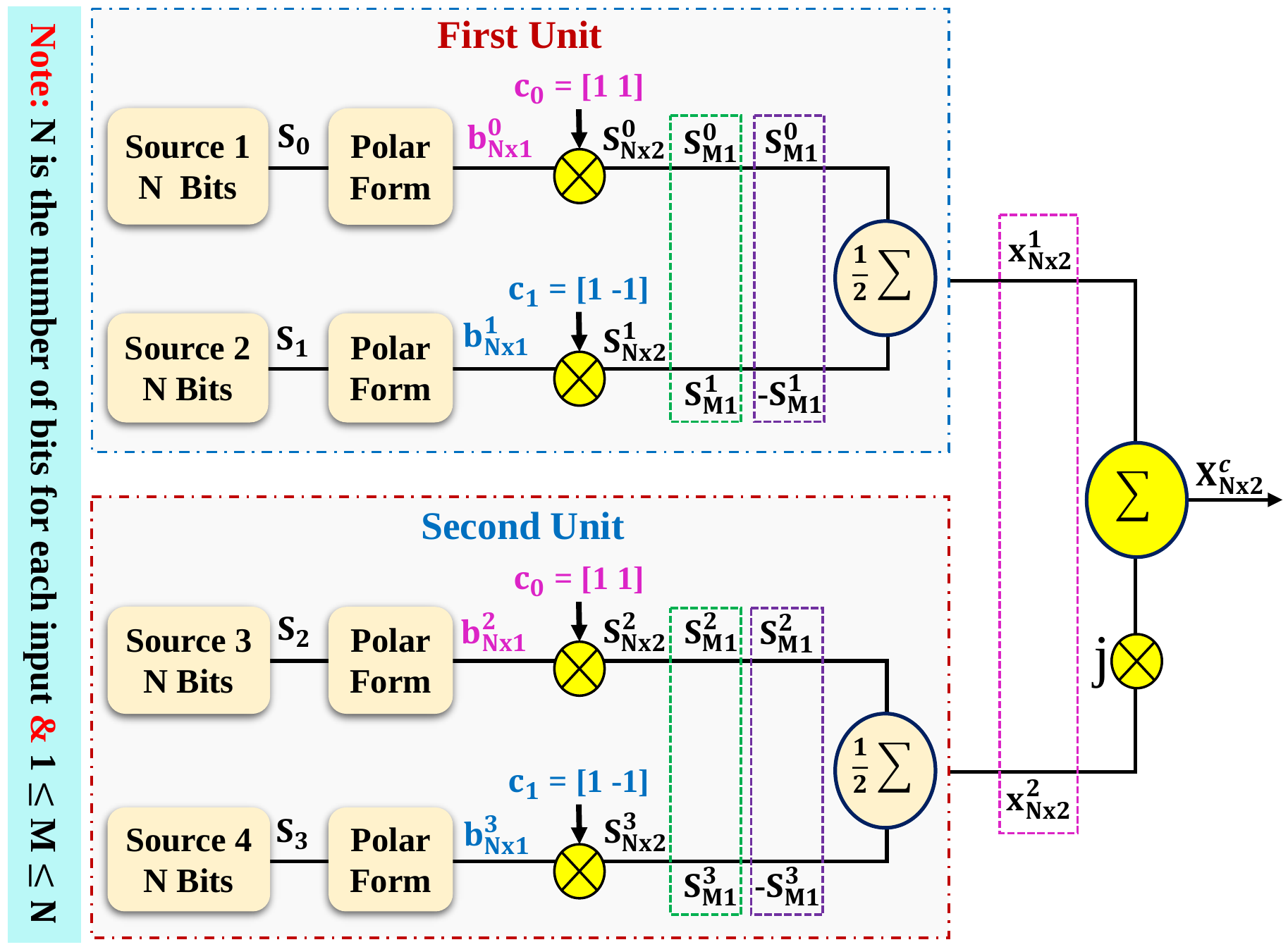}
	\caption{\label{M_STC} Block diagram of M-STC technique.}	
\end{figure}

On the transmitter side, the M-STC scheme is applied through two procedures. The spreading procedure is carried out first, and subsequently the combining process. Two comparable units are joined to create the M-STC technique. The output of the second unit is multiplied by "$j$" to obtain the imaginary component, which is then added to the output of the first unit to produce the complex output $X_c$, as depicted in Fig. \ref{M_STC}.

Fig. \ref{M_STC} displays the block diagram of M-STC technique. First, the input data ($S_0$, $S_1$, $S_2$ and $S_3$) is converted to polar form ($\mathrm {b_{Nx1}^0}$, $\mathrm {b_{Nx1}^1}$, $\mathrm {b_{Nx1}^2}$ and $\mathrm {b_{Nx1}^3}$). The polar form for the first unit is ($\mathrm {b_{Nx1}^0}$ and $\mathrm {b_{Nx1}^1}$) and for the second unit is ($\mathrm {b_{Nx1}^2}$ and $\mathrm {b_{Nx1}^3}$). The polar form for the two units is expressed as follows: 
\begin{equation}
\begin{aligned}
b_{N \times 1}^{0} &=\left[\begin{array}{l}
b_{11}^0 \\
b_{21}^0 \\
\hspace{1.5 mm}.\\
b_{M1}^0 \\
\hspace{3 mm}.\\
b_{N 1}^0
\end{array}\right] \hspace{1.5 cm}
b_{N \times 1}^{1} &=\left[\begin{array}{l}
b_{11}^1 \\
b_{21}^1 \\
\hspace{1.5 mm}.\\
b_{M1}^1 \\
\hspace{3 mm}.\\
b_{N 1}^1
\end{array}\right]\\\\
b_{N \times 1}^{2} &=\left[\begin{array}{l}
b_{11}^2 \\
b_{21}^2 \\
\hspace{1.5 mm}.\\
b_{M1}^2 \\
\hspace{3 mm}.\\
b_{N1}^2
\end{array}\right] \hspace{1.5 cm}
b_{N \times 1}^{3} &=\left[\begin{array}{l}
b_{11}^3 \\
b_{21}^3 \\
\hspace{1.5 mm}.\\
b_{M1}^3 \\
\hspace{3 mm}.\\
b_{N1}^3
\end{array}\right]
\end{aligned}
\end{equation}

Second, the polar form is spread using the Walsh code ($c$), which is constructed using the Hadamard matrix ($H$). The Hadamard matrix is a symmetric square matrix, and each row of the Hadamard matrix is orthogonal to every other row. In this paper, the (2x2) Hadamard matrix is used and it is given as follows \cite{goldsmith2005wireless}:
\begin{equation}
\mathrm {H_{2x2}} =\left[\begin{array}{ll}
1 &\quad 1 \\
1 & \hspace{1 mm}-1 \\
\end{array}\right]
\end{equation}

The Hadamard matrix represents a different Walsh code and each row or column of this matrix represents a different Walsh code. The two spreading Walsh codes are given as follows \cite{el2017symbol, elbakry2022throughput}:
\begin{equation}
c_0 =[1 \quad  1]   \quad and \quad c_1=[1 \quad -1].
\end{equation}

The spread data is obtained by multiplying the polar data ($\mathrm {b_{Nx1}^0}$, $\mathrm {b_{Nx1}^1}$, $\mathrm {b_{Nx1}^2}$ and $\mathrm {b_{Nx1}^3}$) by the Walsh codes ($c_0$ and $c_1$) as follows:

\begin{equation}
\resizebox{.9\hsize}{!}
{$
\begin{aligned}
S_{N \times 2}^{0} &=b_{N \times 1}^{0} \times c_0=\left[\begin{array}{ll}
S_{11}^0 & S_{12}^0\\
S_{21}^0 & S_{22}^0\\
\hspace{0.9 cm}.\\
S_{M1}^0 & S_{M2}^0\\
\hspace{0.9 cm}.\\
S_{N 1}^0 & S_{N 2}^0 
\end{array}\right] = \left[\begin{array}{ll}
b_{11}^0 & \hspace{3 mm} b_{11}^0\\
b_{21}^0 & \hspace{3 mm} b_{21}^0\\
\hspace{0.9 cm}.\\
b_{M1}^0 & \hspace{3 mm} b_{M1}^0\\
\hspace{0.9 cm}.\\
b_{N 1}^0 & \hspace{3 mm} b_{N 1}^0 
\end{array}\right]\\ \\
S_{N \times 2}^{1} &=b_{N \times 1}^{1} \times c_1=\left[\begin{array}{ll}
S_{11}^1 & S_{12}^1\\
S_{21}^1 & S_{22}^1\\
\hspace{0.9 cm}.\\
S_{M1}^1 & S_{M2}^1\\
\hspace{0.9 cm}.\\
S_{N 1}^1 & S_{N 2}^1 
\end{array}\right] = \left[\begin{array}{ll}
b_{11}^1 & -b_{11}^1\\
b_{21}^1 & -b_{21}^1\\
\hspace{0.9 cm}.\\
b_{M1}^1 & -b_{M1}^1\\
\hspace{0.9 cm}.\\
b_{N 1}^1 & -b_{N 1}^1 
\end{array}\right]\\ \\
\end{aligned}$}
\end{equation}

\begin{equation}
\resizebox{.9\hsize}{!}
{$
\begin{aligned}
S_{N \times 2}^{2} &=b_{N \times 1}^{2} \times c_0=\left[\begin{array}{ll}
S_{11}^2 & S_{12}^2\\
S_{21}^2 & S_{22}^2\\
\hspace{0.9 cm}.\\
S_{M1}^2 & S_{M2}^2\\
\hspace{0.9 cm}.\\
S_{N1}^2 & S_{N2}^2 
\end{array}\right] = \left[\begin{array}{ll}
b_{11}^2 & \hspace{3 mm} b_{11}^2\\
b_{21}^2 & \hspace{3 mm} b_{21}^2\\
\hspace{0.9 cm}.\\
b_{M1}^2 & \hspace{3 mm} b_{M1}^2\\
\hspace{0.9 cm}.\\
b_{N1}^2 & \hspace{3 mm} b_{N1}^2 
\end{array}\right]\\ \\
S_{N \times 2}^{3} &=b_{N \times 1}^{3} \times c_1=\left[\begin{array}{ll}
S_{11}^3 & S_{12}^3\\
S_{21}^3 & S_{22}^3\\
\hspace{0.9 cm}.\\
S_{M1}^3 & S_{M2}^3\\
\hspace{0.9 cm}.\\
S_{N1}^3 & S_{N2}^3 
\end{array}\right] = \left[\begin{array}{ll}
b_{11}^3 & -b_{11}^3\\
b_{21}^3 & -b_{21}^3\\
\hspace{0.9 cm}.\\
b_{M1}^3 & -b_{M1}^3\\
\hspace{0.9 cm}.\\
b_{N1}^3 & -b_{N1}^3 
\end{array}\right].\\
\end{aligned}$}
\end{equation}

Where $S_{N \times 2}^{0}$  and $S_{N \times 2}^{1}$ are the spread data of the first unit, whereas $S_{N \times 2}^{2}$ and $S_{N \times 2}^{3}$ are the spread data of the second unit. The combining process is then done to the spread data for both the first and second unit after the spreading procedure. To obtain the combining data $\mathrm{{x}_{Nx2}^1}$, the spread data $S_{N \times 2}^{0}$  and $S_{N \times 2}^{1}$ are merged in the first unit. Similarly, the spread data $S_{N \times 2}^{2}$  and $S_{N \times 2}^{3}$ are merged in the second unit to get the combining data $\mathrm{{x}_{Nx2}^2}$. The combining process for first and second unit is given as follows: 
\begin{equation}
\begin{aligned}
\mathrm{{x}_{Nx2}^1}
& =\frac{1}{2} \left[ \left( S_{M1}^0+S_{M1}^1\right) \quad \left( S_{M1}^0-S_{M1}^1\right)\right].\\
\mathrm{{x}_{Nx2}^2} 
& =\frac{1}{2} \left[ \left( S_{M1}^2+S_{M1}^3\right) \quad \left( S_{M1}^2-S_{M1}^3\right)\right]. 
\end{aligned}
\end{equation}

Where, $\mathrm{{x}_{Nx2}^1}$ is the combining data for the first unit, $\mathrm{{x}_{Nx2}^2}$ is the combining data for the second unit, $ 1 \leq M \leq N$ and $N$ represents the input's bits. As displayed in Fig. \ref{M_STC}, the complex output $X_c$ is created by adding the imaginary component, obtained by multiplying the output of the second unit by "j," to the first unit's output. The output of the M-STC technique is expressed as follows: 
\begin{equation}
\mathrm{X_{Nx2}^c} = \mathrm{{x}_{Nx2}^1} + j\mathrm{{x}_{Nx2}^2}.
\end{equation}

\begin{figure*}[!ht]
	\centering
	\includegraphics[width=0.8\linewidth,height=8. cm]{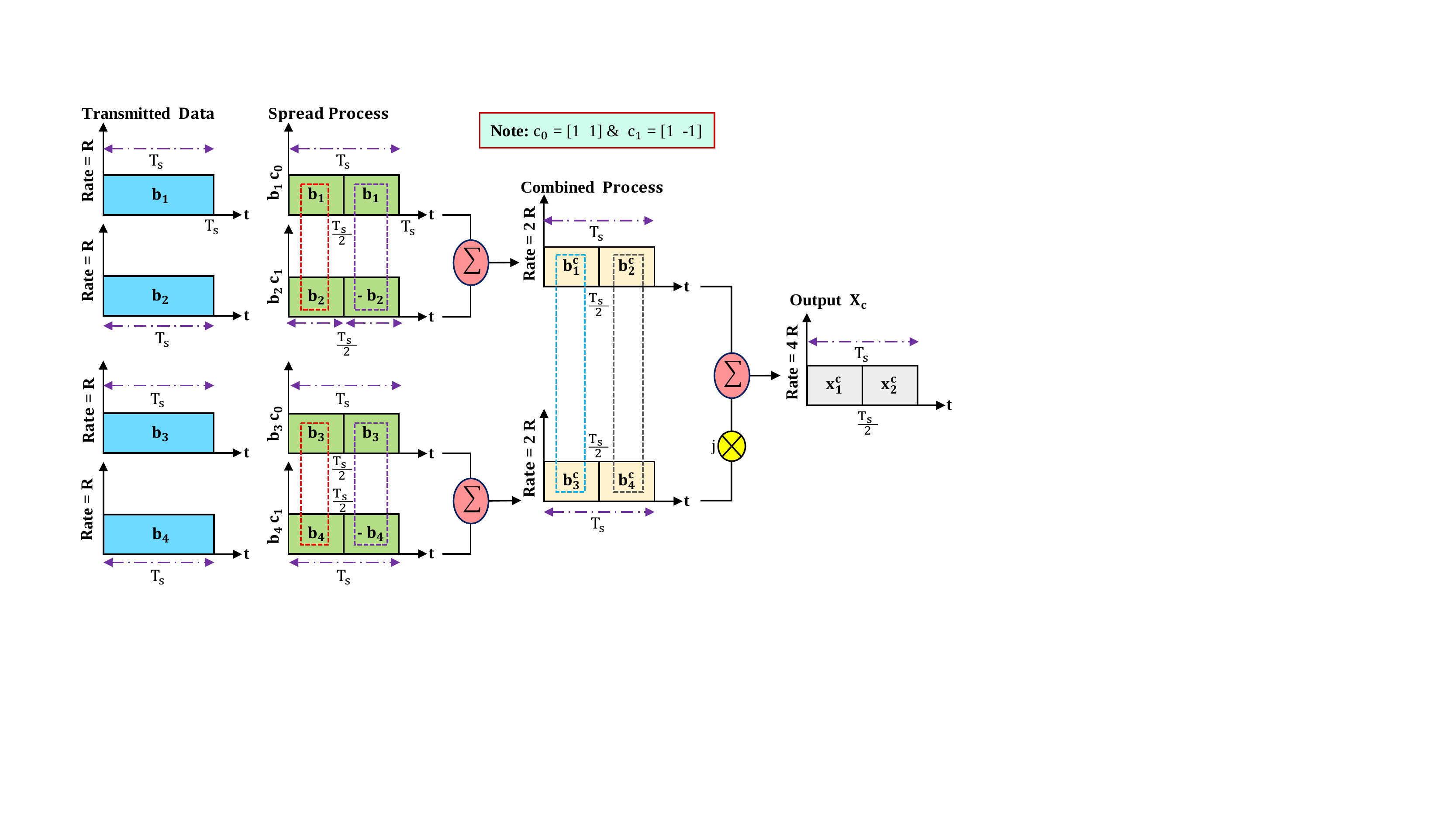}
	\caption{\label{Waveforms_of_M_STC} Illustrative example of the increasing data rate using the M-STC technique.}	
\end{figure*}

Fig. \ref{Waveforms_of_M_STC} shows an illustrative  example to explain how the M-STC technique employs the spread and combining process to compress the symbol time and enhance the data rate. Instead of sending one bit per symbol time, the M-STC technique sends four bits per symbol time. However, as will be shown in section V, the BER obtained using the M-STC approach is the same as the BER obtained using BPSK modulation. In Fig. \ref{Waveforms_of_M_STC}, we transmit 4 bits ($b_1$, $b_2$, $b_3$ and $b_4$), where each transmitted bit has a rate $R$. The transmitted bits $b_1$ and $b_2$ are multiplied by Walsh codes $c_0$ and $c_1$ ($c_0 = [1\quad 1]$ and $c_1 = [1\quad -1]$) respectively to obtain spread data as follows:
\begin{equation}
\begin{aligned}
S_{d1} &= b_1 \times c_0 = [b_1 \hspace{7.6 mm} b_1]\\
S_{d2} &= b_2 \times c_1 = [b_2 \quad -b_2],
\end{aligned}
\end{equation}
where $S_{d1}$ and $S_{d2}$ are spread data for the transmitted data $b_1$ and $b_2$ respectively. Similarly, for the transmitted data $b_3$ and $b_4$, the spread data is given as follows:
\begin{equation}
\begin{aligned}
S_{d3} &= b_3 \times c_0 = [b_3 \hspace{7.6 mm} b_3]\\
S_{d4} &= b_4 \times c_1 = [b_4 \quad -b_4].
\end{aligned}
\end{equation}
Where $S_{d3}$ and $S_{d4}$ are spread data for the transmitted data $b_3$ and $b_4$ respectively. The combining process for spread data $S_{d1}$ and $S_{d2}$ is defined as follows:
\begin{equation}
\begin{aligned}
C_{d1} & =  [b_1^c \hspace{7.6 mm} b_2^c]\\
b_1^c & = b_1 + b_2\\
b_2^c & = b_1 - b_2.
\end{aligned}
\end{equation}

Similarly for the combining process for spread data $S_{d3}$ and $S_{d4}$ is expressed as follows:
\begin{equation}
\begin{aligned}
C_{d2} & =  [b_3^c \hspace{7.6 mm} b_4^c]\\
b_3^c & = b_3 + b_4\\
b_4^c & = b_3 - b_4.
\end{aligned}
\end{equation}

It is clear from Fig. \ref{Waveforms_of_M_STC} that the rate of combining data $C_{d1}$ = $2R$ and the rate of $C_{d2}$ = $2R$. The output of the M-STC technique  is written as follows:
\begin{equation}
\begin{aligned}
X_c & =  [\mathrm{x_1^c} \hspace{7.6 mm} \mathrm {x_2^c}]\\
\mathrm{x_1^c} & = b_1^c + j b_3^c\\
\mathrm{x_2^c} & =  b_2^c + j b_4^c.
\end{aligned}
\end{equation}

Where, $X_c$ is the output of the M-STC technique. The output $X_c$ has a data rate of 4R. Therefore, employing the M-STC technique results in increasing in capacity by compressing the symbol time and transferring data at a higher rate.
 
\subsection{M-STE System Model}
\begin{figure*}[!ht]
	\centering
	\includegraphics[width=0.9\linewidth,height= 6.2 cm]{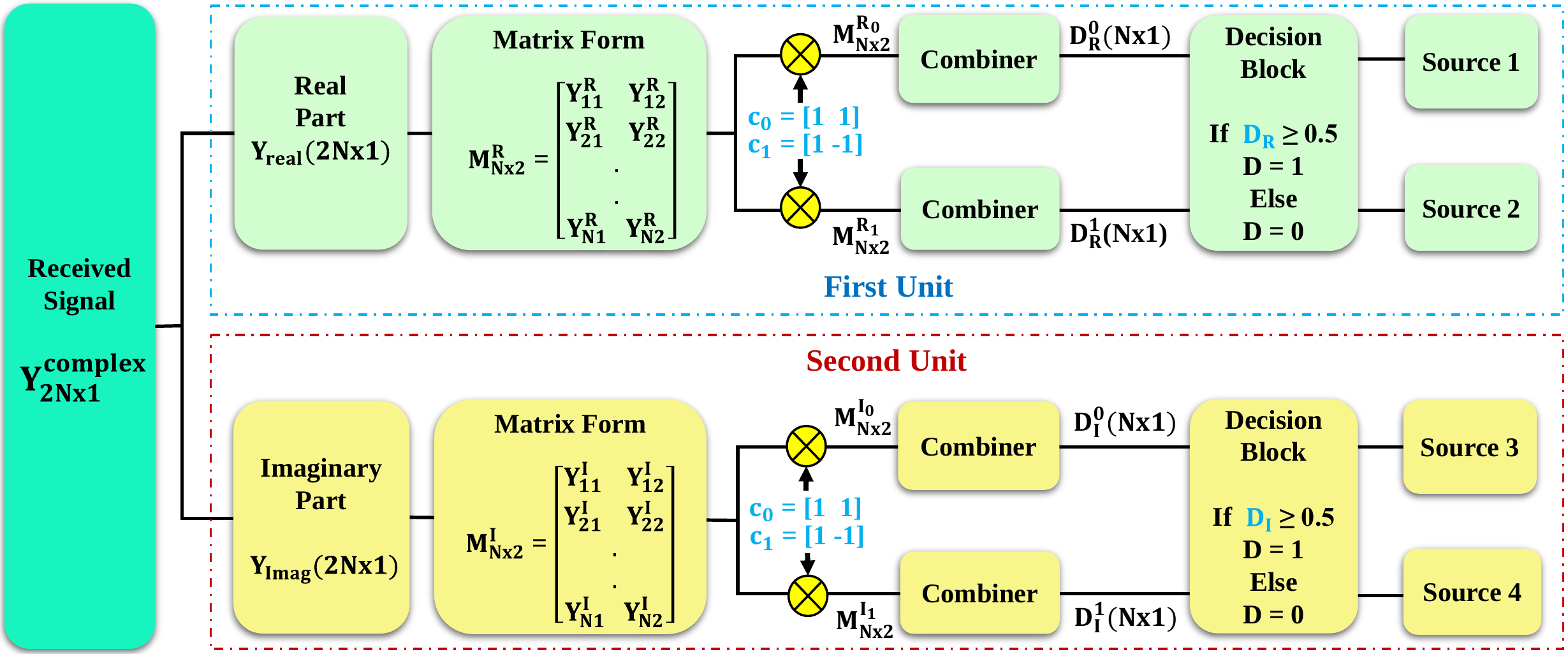}
	\caption{\label{M_STE} Block diagram of M-STE technique.}	
\end{figure*}

The mathematical analysis is explained in details for the M-STE technique in this subsection. In order to reverse the procedures that the M-STC technique performed in the transmitter, the M-STE technique is employed at the receiver side. As indicated in Fig. \ref{M_STE}, the received signal ($\mathrm{Y_{2Nx1}^c} = \mathrm{{Y}_{real} + j\mathrm{Y}_{imag}}$) is divided into two units: real (Unit 1) and imaginary (Unit 2). For the first unit, the real portion of the received signal is initially transformed to a $N$x$2$ matrix in the following manner:
\begin{equation}
\label{Matx_R}
\mathrm {M_{N \times 2}^R=\left[\begin{array}{ll}
Y_{11}^R & Y_{12}^R \\
Y_{21}^R & Y_{22}^R \\
& \cdot \\
& \cdot \\
Y_{N 1}^R & Y_{N2}^R
\end{array}\right]}.
\end{equation}

The Walsh codes  $c_0$ and $c_1$ are multiplied by equation (\ref{Matx_R}) to disseminate data in the first unit as follows:
\begin{equation}
\begin{aligned}
\mathrm {M_{N \times 2}^{R_0}=M_{N \times 2}^R \times c_0} &=\left[\begin{array}{ll}
Y_{11}^R & \hspace{3.5 mm} Y_{12}^R \\
Y_{21}^R & \hspace{3.5 mm} Y_{22}^R \\
& \cdot \\
& \cdot \\
Y_{N 1}^R & \hspace{3 mm}Y_{N2}^R
\end{array}\right]\\ \\
\mathrm {M_{N \times 2}^{R_1}=M_{N \times 2}^R \times c_1} &=\left[\begin{array}{ll}
Y_{11}^R & -Y_{12}^R \\
Y_{21}^R & -Y_{22}^R \\
& \cdot \\
& \cdot \\
Y_{N 1}^R & -Y_{N2}^R
\end{array}\right].
\end{aligned}
\end{equation}

Where, c$_0$=[1 0 ; 0 1] and c$_1$=[1 0 ; 0 -1]. To create the combined data, the combining procedure is used on the spread data. The procedure for combining data is expressed by the equation below:
\begin{equation}
\begin{aligned}
\mathrm {\underbrace{D_R^0}_{Nx1}} &=\frac{ \sum_{i=1}^{N}\left(M_{i 1}^{R_0} + M_{i 2}^{R_0}\right) +1}{2} \\
\mathrm {\underbrace{D_R^1}_{Nx1}} &=\frac{ \sum_{i=1}^{N}\left(M_{i 1}^{R_1} + M_{i 2}^{R_1}\right) +1}{2}.
\end{aligned}
\end{equation}

Similar to the real part (Unit 1), the second unit repeats all of the previous unit's steps. The imaginary portion of the received signal is therefore expressed as follows:
\begin{equation}
\label{Matx_I}
\mathrm {M_{N \times 2}^I=\left[\begin{array}{ll}
	Y_{11}^I & Y_{12}^I \\
	Y_{21}^I & Y_{22}^I \\
	& \cdot \\
	& \cdot \\
	Y_{N 1}^I & Y_{N2}^I
	\end{array}\right]}
\end{equation}

The equation (\ref{Matx_I}) is multiplied by Walsh codes  $c_0$ and $c_1$ to obtain the spread data in the second unit:
\begin{equation}
\begin{aligned}
\label{spread_I}
\mathrm {M_{N \times 2}^{I_0}=M_{N \times 2}^I \times c_0} &=\left[\begin{array}{ll}
Y_{11}^I & \hspace{3.5 mm} Y_{12}^I \\
Y_{21}^I & \hspace{3.5 mm} Y_{22}^I \\
& \cdot \\
& \cdot \\
Y_{N 1}^I & \hspace{3 mm}Y_{N2}^I
\end{array}\right]\\ \\
\mathrm {M_{N \times 2}^{I_1}=M_{N \times 2}^I \times c_1} &=\left[\begin{array}{ll}
Y_{11}^I & -Y_{12}^I \\
Y_{21}^I & -Y_{22}^I \\
& \cdot \\
& \cdot \\
Y_{N 1}^I & -Y_{N2}^I
\end{array}\right].
\end{aligned}
\end{equation}

The spread data in equation (\ref{spread_I}) is merged as follows to generate the combined data in the second unit:
\begin{equation}
\begin{aligned}
\mathrm {\underbrace{D_I^0}_{Nx1}} &=\frac{ \sum_{i=1}^{N}\left(M_{i 1}^{I_0} + M_{i 2}^{I_0}\right) +1}{2} \\
\mathrm {\underbrace{D_I^1}_{Nx1}} &=\frac{ \sum_{i=1}^{N}\left(M_{i 1}^{I_1} + M_{i 2}^{I_1}\right) +1}{2}.\\
\end{aligned}
\end{equation}

Finally, the combined data is inserted to the decision block in order to recover the transmitted data.
\subsection{M-STC-OFDM System Model}
\begin{figure}[!ht]
	\centering
	\includegraphics[width=0.95\linewidth,height= 4.9 cm]{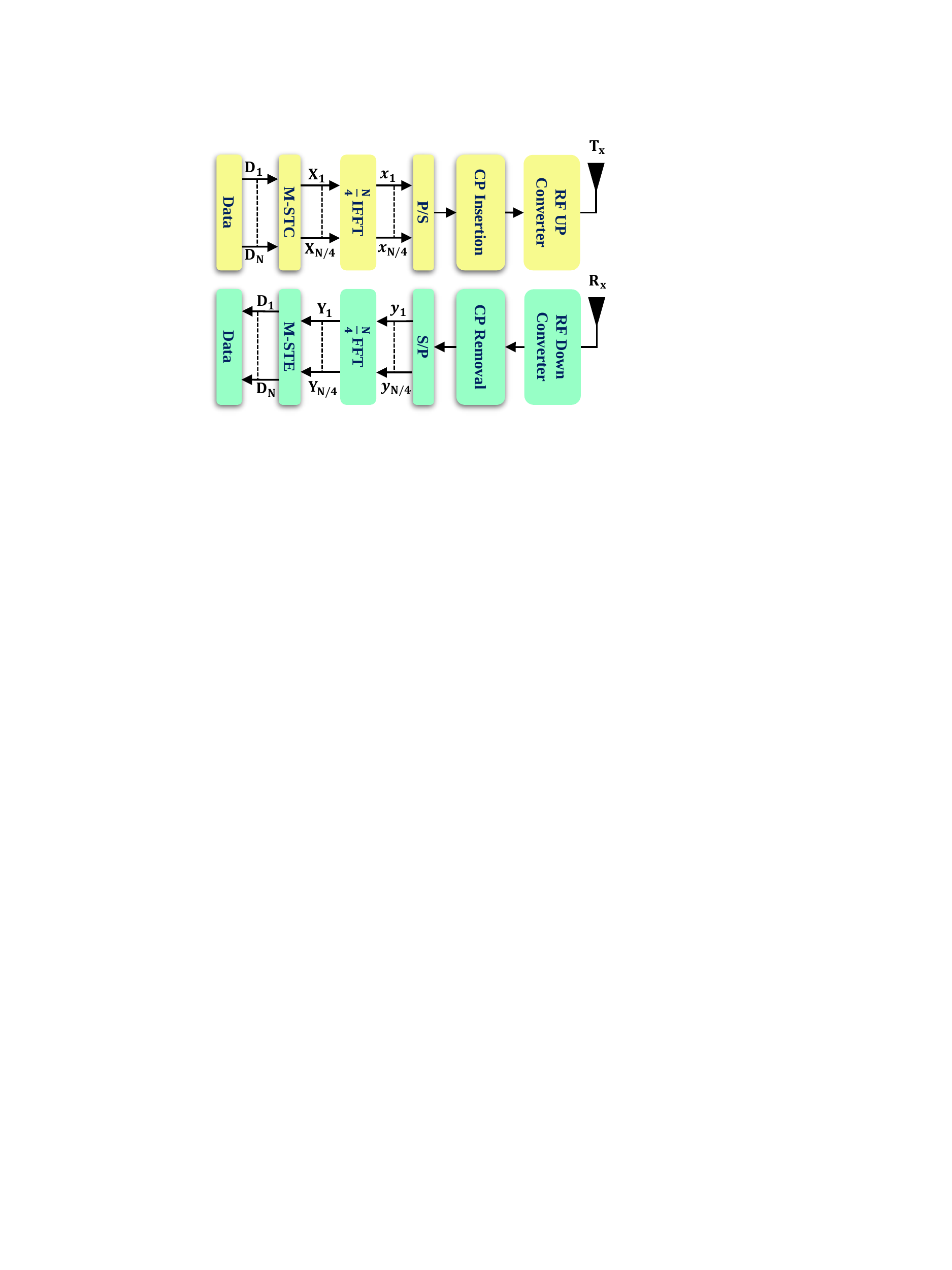}
	\caption{\label{Block_Diag_OFDM_M_STC} Block diagram of M-STC-OFDM system.}	
\end{figure}

The mathematical analysis of the OFDM system using the M-STC technique (M-STC-OFDM) is presented in this subsection. Fig. \ref{Block_Diag_OFDM_M_STC} depicts  a generic block diagram of an OFDM transceiver sequence that uses the M-STC at the transmitter and the M-STE at the receiver to compress symbol time, boost data rate, and minimize PAPR. To reduce the symbol time to one-fourth of its original length and increase capacity, the input data ($D_0, D_1,..., D_N$) is first processed via the M-STC block. Therefore, rather than using one bit for each symbol, four bits are used instead. As a consequence, the M-STC block's output ranges from $X_0$ to $X_{N/4}$. The complex data symbol on the $K^{th}$ sub-carrier is denoted by $X_k$, $k=1,2,..., N/4$. The $N/4$ resultant  wave-forms are transmitted into the $N/4$ input ports of an inverse fast Fourier transform (IFFT) block. Following IFFT, a discrete-time OFDM symbol is represented in the form: 
\begin{equation}
x_{k} =\frac{2}{N} \sum_{m=0}^{\frac{N}{4}-1} X_{m} e^{j 2 \pi k m /\frac{N}{4}}, \quad 0 \leq k \leq \frac{N}{4}-1
\end{equation}

Where $k$ indicates the time index, $N$ is indeed  the number of sub-carriers, $x_{k}$ is in fact  the $k^{th}$ OFDM symbol and $X_m$ denotes the  $m^{th}$ transferred data symbols. The generated time domain symbols are passed through a parallel-to-serial (P/S) converter. To ensure the orthogonality and avoid ICI and ISI, the cyclic prefix (CP) is placed before each OFDM signal as a guard interval (GI) between OFDM symbols. A  CP of a appropriate length ($L_{cp}$) is applied to mitigate the impact of multi-path propagation and the transmitted OFDM symbol with CP is written as follows:
\begin{equation}
x_{k}^{cp} =\frac{2}{N} \sum_{m=0}^{\frac{N}{4}-1} X_{m} e^{j 2 \pi k m /\frac{N}{4}},  -L_{cp} \leq k \leq \frac{N}{4}-1
\end{equation}

As indicated in Fig. \ref{Block_Diag_OFDM_M_STC}, the transmitter procedures is effectively reversed in opposite order at the receiver side in order to retrieve the sent data.
\section{Peak-to-Average-Power-Ratio (PAPR)}
\begin{figure}[!ht]
	\centering
	\includegraphics[width=0.95\linewidth,height= 6.6 cm]{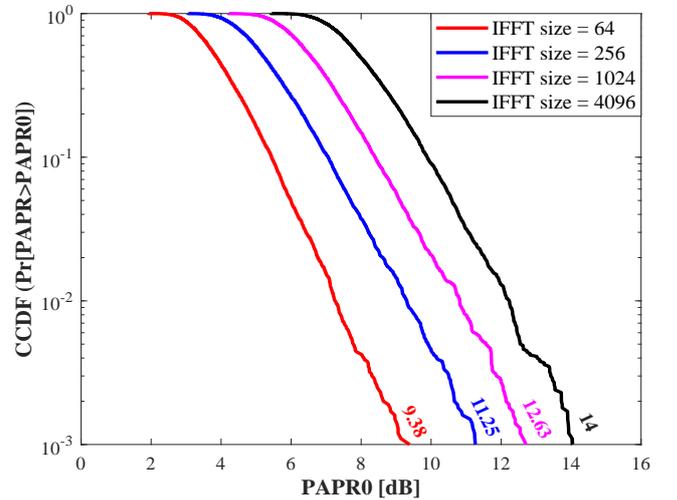}
	\caption{\label{IFFT_paper} CCDF of PAPR for an OFDM signal with different IFFT size.}	
\end{figure}

The main disadvantage of the OFDM system is high PAPR. The high PAPR is caused by the multi-carrier of the OFDM signal. When the $N$ signals in an OFDM system add up to the same phase, the PAPR increases significantly \cite{mohammed2021novel}. The measured PAPR may reach N times the average OFDM symbol amplitude \cite{mohammed2018peak}. The PAPR of an OFDM signal $x(n)$ is defined as the ratio of its peak power to its average power.  The PAPR in [dB] is mathematically defined as follows \cite{mohammed2019performance}:  
\begin{align}
PAPR\{x(n)\}& =  \dfrac{P_{peak}}{P_{avg}} =   10.\log_{10} \dfrac{\max{|x(n)|^2}}{\textbf{E}\{|x(n)|^2\}}
\end{align}

where, $\textbf{E}\{.\}$  is the expectation operator, $P_{peak}$ is the peak power and $P_{avg}$ is the average power of the OFDM signal. Fig. \ref{IFFT_paper} displays the CCDF of the PAPR for an OFDM signal assuming BPSK modulation and various IFFT sizes. The relationship between growing IFFT size and rising the PAPR is shown in Fig. \ref{IFFT_paper}.

\section{Simulation Results and Discussion}
This section provides numerical simulation results for the OFDM system that use the M-STC/M-STE techniques mentioned in Section III, with performance measures such as PAPR, BER, OFDM symbol time, and PSD of interest. The system's simulation parameters are shown in Table \ref{parametter}.
\renewcommand{\arraystretch}{1.2}
\small\addtolength{\tabcolsep}{-1. pt}
\begin{table}[htbp]
	\centering
	\caption{Simulation Parameters \cite{DoublingIoT}}
	\begin{tabular}{|l|l|}
		\hline
		\textbf{Parameter }
		&  \textbf{{Value and Unit}} 
		\\
		\hline
		\hline
		
		Occupied Channel Bandwidth
		&180 kHz\\		
		Spacing frequency, $\varDelta$f
		&15 kHz \\
		Sampling frequency , $f_s$
		& 1.92 MHz \\
		FFT size, $N$
		& 128 \\
		CP length
		& 1/4 of the OFDM symbol \\
		Modulation type
		& BPSK \\
		Channel model
		& AWGN \\
		\hline
	\end{tabular}%
	\label{parametter}%
\end{table}

\begin{figure}[!ht]
	\centering
	\includegraphics[width= 1.00\linewidth,height=10.0 cm]{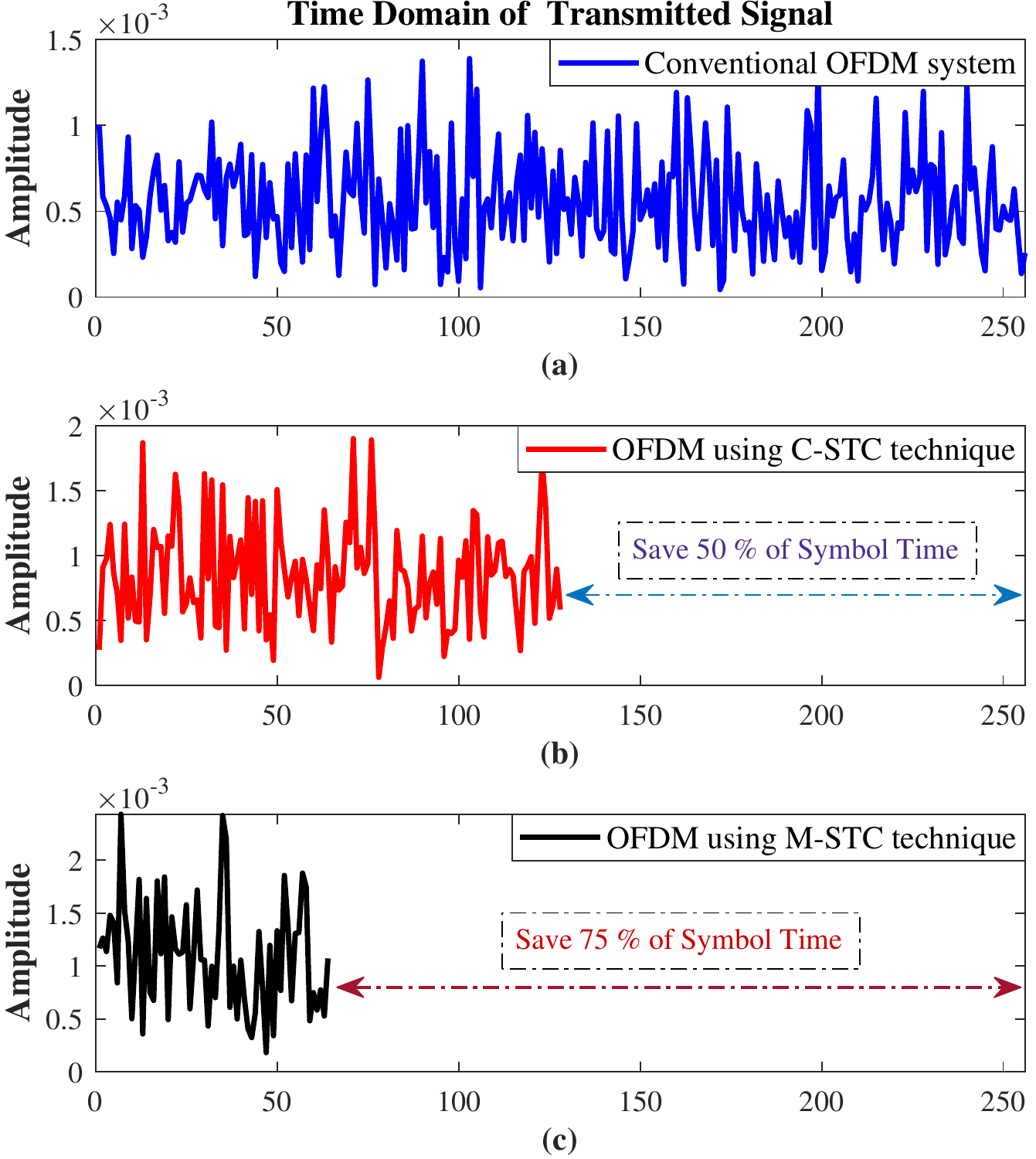}
	\caption{Performance comparison between time domain of the transmitted signal: (a) conventional  OFDM system, (b) C-STC-OFDM system \cite{afifi2020efficient} and (c) M-STC-OFDM system.}
	\label{Time_Domain}
\end{figure}

In the OFDM systems, the time domain of the transmitted signal is measured after IFFT in the transmitter side. Fig. \ref{Time_Domain} shows the performance comparison between time domain of typical OFDM, C-STC-OFDM, and M-STC-OFDM system. As seen in Fig. \ref{Time_Domain} (b), using C-STC-OFDM in \cite{afifi2020efficient} reduces the time required to process an OFDM symbol by 50\% when compared to a conventional OFDM system. While, using the suggested approach (M-STC-OFDM) significantly reduces the time required for each OFDM symbol by 75\% when compared to conventional OFDM system and 50\% when compared to C-STC-OFDM system, as it is clear in Fig. \ref{Time_Domain} (c).

\begin{figure}[!ht]
	\centering
	\includegraphics[width=0.95 \linewidth,height=16.2 cm]{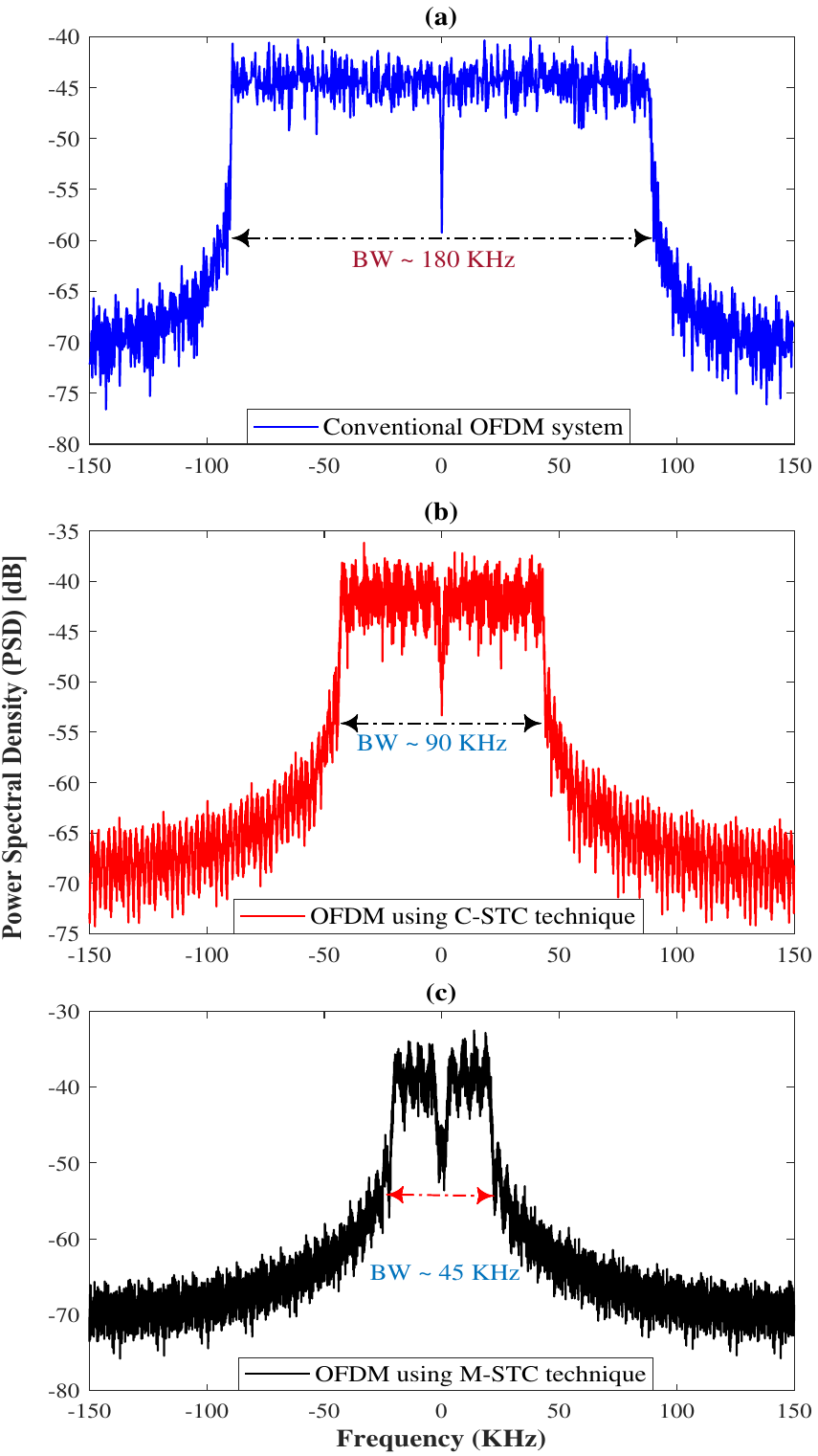}
	\caption{PSD-based performance comparison between: (a) conventional  OFDM system, (b) C-STC-OFDM system and (c) M-STC-OFDM system.}
	\label{PSD}
\end{figure}

The spectrum for typical OFDM, which has a 180 kHz bandwidth, can be seen in the Fig. \ref{PSD} (a), while the spectrum for the C-STC-OFDM system, that has a 90 kHz bandwidth, is shown in Fig. \ref{PSD} (b).  As it is clearly seen in Fig. \ref{PSD} (c), the third spectrum is for the M-STC-OFDM system, whose bandwidth is 45 kHz. The C-STC OFDM system requires half the bandwidth (50\%) in comparison to a typical OFDM system to transfer the same amount of data, as illustrated by the three spectra in Fig. \ref{PSD}. When compared to the standard OFDM system and the C-STC-OFDM system, the M-STC-OFDM needs just a fourth (25\%) and half (50\%) of the bandwidth, respectively, to deliver the same amount of data.

\begin{figure}[!ht]
	\centering
	\includegraphics[width=0.95\linewidth,height= 6.5 cm]{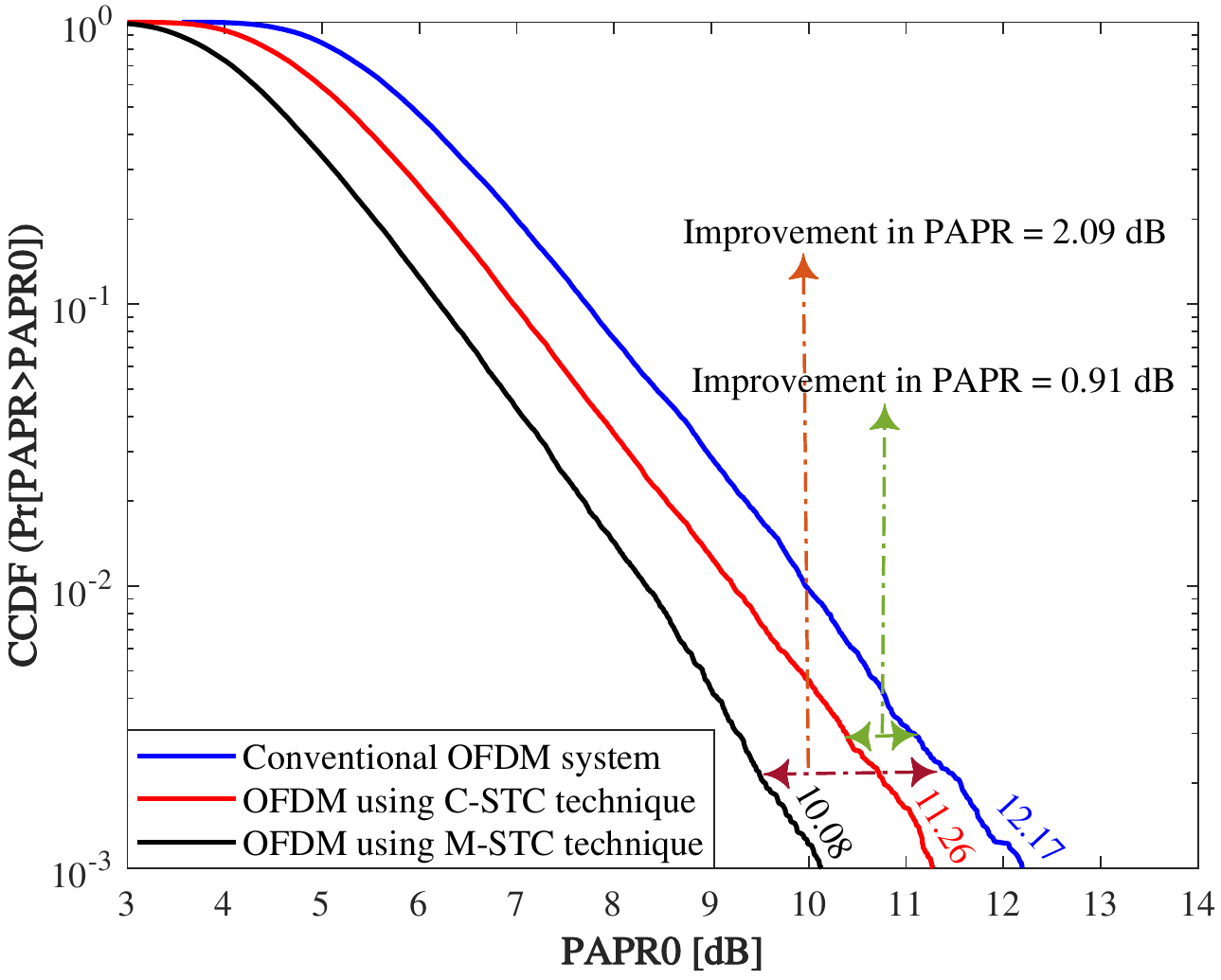}
	\caption{\label{PAPR_3_paper_2} PAPR-based performance comparison between the typical OFDM system, C-STC-OFDM system, and M-STC-OFDM system.}	
\end{figure} 

When the C-STC approach is applied to an OFDM system, the number of IFFT is decreased by half, furthermore the PAPR is also reduced. Implementing the C-STC-OFDM system leads to reduce the PAPR by 0.91 dB, where the PAPR of conventional OFDM is 12.17 dB, whereas the PAPR of C-STC-OFDM is 11.26, as seen in. Fig. \ref{PAPR_3_paper_2}. Applying M-STC technique to an OFDM system leads to reduce the IFFT size to a quarter. Consequently, the PAPR is reduced as well. According to Fig. \ref{PAPR_3_paper_2}, the enhancement in PAPR achieved by the M-STC-OFDM system is 2.09 dB. This means that the PAPR enhancement is 1.18 dB higher than it was with the C-STC-OFDM system, or about 230\% higher.

\begin{figure}[!ht]
	\centering
	\includegraphics[width=0.95\linewidth,height= 6.2 cm]{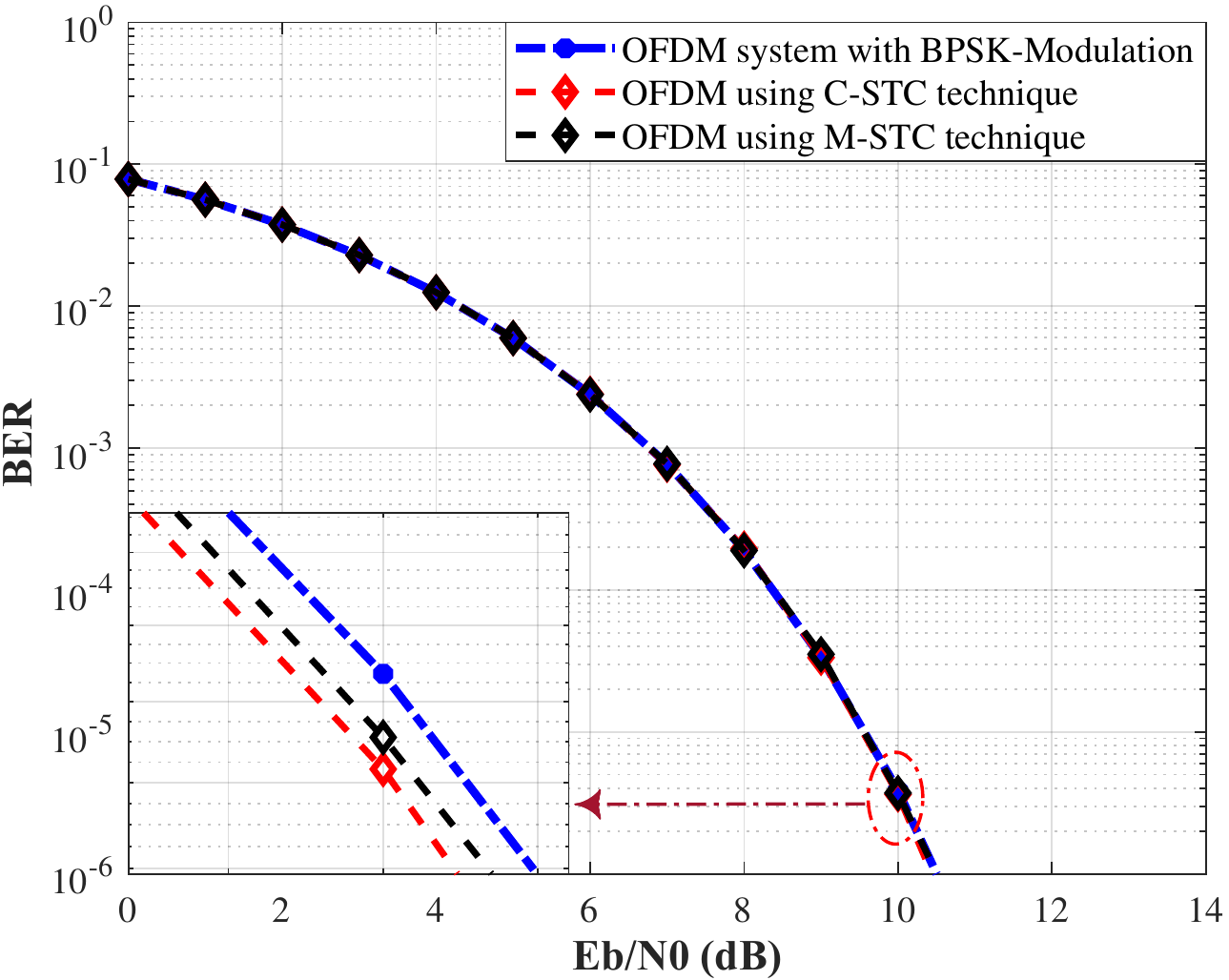}
	\caption{\label{BER_paper_2} BER-based performance comparison between the typical OFDM system, C-STC-OFDM, and  M-STC-OFDM.}	
\end{figure}

Fig. \ref{BER_paper_2} compares the BER performance of a typical OFDM system, the C-STC-OFDM system, and the M-STC-OFDM system. It is observed that the three systems (typical OFDM, C-STC-OFDM and M-STC-OFDM) have the same BER. When the M-STC scheme is added to an OFDM system to create an M-STC-OFDM, 75\% of the OFDM symbol time is saved and the bandwidth is reduced to a quarter (75\% of BW is saved). As previously depicted in Figs. \ref{Time_Domain}, \ref{PSD} and \ref{PAPR_3_paper_2}, the M-STC-OFDM system outperforms the C-STC-OFDM system in terms of bandwidth by 50\%, OFDM symbol time by 50\%, and PAPR by 1.18 dB (about 230\%). Nevertheless,  the proposed M-STC-OFDM system has the same BER as standard OFDM and C-STC-OFDM systems. 

\begin{figure}[!ht]
	\centering
	\includegraphics[width= 0.95\linewidth,height = 6.2 cm]{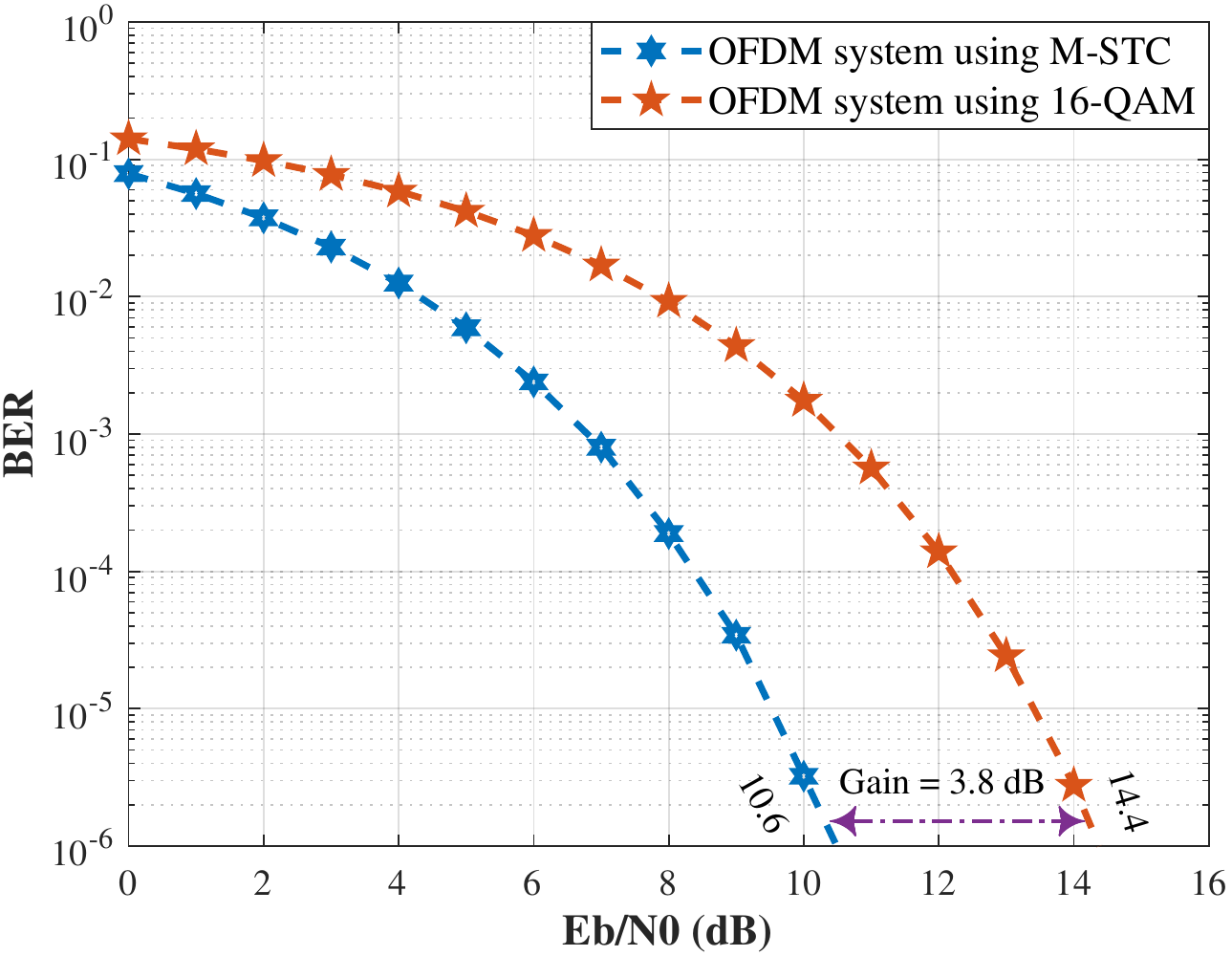}
	\caption{BER based performance comparison between the M-STC-OFDM system and a typical OFDM system using 16-QAM}
	\label{BER_QAM_paper_2}
\end{figure}

Fig. \ref{BER_QAM_paper_2} displays the BER of the proposed system (M-STC-OFDM) and the typical OFDM system using 16-QAM modulation. At the same BER ($BER = 10^{-6}$) and for the same amount of data, the signal to noise ratio (SNR) of the typical OFDM system with 16-QAM modulation is 14.4 dB, whereas the SNR of the M-STC-OFDM system is 10.6 dB. Therefore, the SNR gain while employing the M-STC-OFDM system is 3.8 dB.  Consequently, it may be concluded that the M-STC-OFDM system has the capacity to transport data at a high rate without consuming more power.
 
\section{Complexity and Algorithms}
The algorithms used in the M-STC-OFDM system are described in this section. The M-STC scheme is represented by Algorithm \ref{M_STC_Tech} and the M-STE approach is illustrated by Algorithm \ref{M_STE_Tech}. The computational complexity of the traditional OFDM system, the C-STC-OFDM system, and the suggested M-STC-OFDM system is seen in the Table \ref{Complex}. The complexity of the suggested system M-STC-OFDM is roughly the same as that of the traditional OFDM system. In other words, all systems have the same order of complexity.
\begin{table}[ht!]
	\renewcommand{\arraystretch}{1.2}
	\centering
	\caption{Computational Complexity Analysis.}
	\resizebox{\columnwidth}{!}{	\begin{tabular}{|c|c|c|}
			\hline
			\textbf{Techniques} & \textbf{ No. Multiplications} & \textbf{ No. Additions} \\
			\hline
			\hline
			Conventional OFDM  \cite{afifi2020efficient}
			& 2N Log$_2$N - 2N 
			& 3N Log$_2$N - N  \\
			\hline
			C-STC-OFDM  \cite{afifi2020efficient}
			& N Log$_2$$\mathrm{\frac{N}{2}}$ - N 
			& $\mathrm{\frac{3}{2}}$N Log$_2$$\mathrm{\frac{N}{2}}$ - $\mathrm{\frac{N}{2}}$   \\
			\hline
			M-STC-OFDM 
			& $\mathrm{\frac{N}{2}}$ Log$_2$$\mathrm{\frac{N}{4}}$ - $\mathrm{\frac{N}{2}}$  
			& $\frac{3}{4}$N Log$_2$$\mathrm{\frac{N}{4}}$ - $\mathrm{\frac{N}{4}}$  \\
			\hline
			\end{tabular}}
	\label{Complex}
\end{table}

\begin{algorithm}[ht!]
	\caption{: The M-STC Technique}
	\label{M_STC_Tech}
	\begin{algorithmic}
		\STATE \textbf{First unit}
	\end{algorithmic}
	\begin{algorithmic}[1]
		\STATE S$_0$ = N Stream of bits from Input 1;
		\STATE S$_1$ = N Stream of bits from Input 2;
		\STATE {b$_{\mathrm{Nx1}}^0$} = Convert N bits of Input 1 to polar form;
		\STATE b$_\mathrm{Nx1}^1$ = Convert N bits of Input 2 to polar form;
	    \end{algorithmic}

    	\begin{algorithmic}
	    \STATE {$------$ {Spread Process For Source 1} $------$}
       \end{algorithmic}
   
	   \begin{algorithmic}[1]
		\STATE A$_\mathrm{Nx1}^1$ = b$_\mathrm{Nx1}^0$.$*$c$_0$(:,1); $\rightarrow$ c$_0$ = [1 \quad 1]; 
		\STATE A$_\mathrm{Nx1}^2$ = b$_\mathrm{Nx1}^0$.$*$c$_0$(:,2); $\rightarrow$ c$_0$ = [1 \quad 1];  	
		\STATE S$_\mathrm{Nx2}^0$ = [A$_\mathrm{Nx1}^1$ \quad,\quad A$_\mathrm{Nx1}^2$]; 	
	    \end{algorithmic}

       \begin{algorithmic}
	    \STATE {$------$ {Spread Process For Source 2} $------$}
       \end{algorithmic}

       \begin{algorithmic}[1]
		\STATE B$_\mathrm{Nx1}^1$ = b$_\mathrm{Nx1}^1$.$*$c$_1$(:,1); $\rightarrow$ c$_1$ = [1 \hspace{1.6 mm} -1];
		\STATE B$_\mathrm{Nx1}^2$ = b$_\mathrm{Nx1}^1$.$*$c$_1$(:,2); $\rightarrow$ c$_1$ = [1 \hspace{1.6 mm} -1];  	
		\STATE S$_\mathrm{Nx2}^1$ = [B$_\mathrm{Nx1}^1$ \quad,\quad B$_\mathrm{Nx1}^2$]; 
		  		 
	   \end{algorithmic}

       \begin{algorithmic}
       \STATE {$---------$ {Combining Process} $---------$}
       \end{algorithmic}

       \begin{algorithmic}[1]
       	\STATE P$_\mathrm{Nx2}^1$ = [A$_\mathrm{Nx1}^1$ \hspace{0.8 mm},\hspace{0.8 mm} B$_\mathrm{Nx1}^1$];
       	\STATE C$_\mathrm{Nx1}^1$ = $\frac{1}{2}$.$*$ sum $\left(P_\mathrm{Nx2}^1 ,2 \right)$ ;$\rightarrow$ Combined step
       	\STATE P$_\mathrm{Nx2}^2$ = [A$_\mathrm{Nx1}^2$ \hspace{0.8 mm},\hspace{0.8 mm} B$_\mathrm{Nx1}^2$]; 
       	\STATE C$_\mathrm{Nx1}^2$ = $\frac{1}{2}$.$*$sum $\left(P_\mathrm{Nx2}^2 ,2 \right)$ ;$\rightarrow$ Combined step
       	
		\STATE x$_\mathrm{Nx2}^1$ =  [C$_\mathrm{Nx1}^1$ \hspace{0.8 mm},\hspace{0.8 mm} C$_\mathrm{Nx1}^2$];$\rightarrow$ The output of unit 1	
	\end{algorithmic}
    \begin{algorithmic}
	\STATE \textbf{Second unit}
   \end{algorithmic}
\begin{algorithmic}[1]
	\STATE S$_2$ = N Stream of bits from Input 3;
	\STATE S$_3$ = N Stream of bits from Input 4;
	\STATE b$_{\mathrm{Nx1}}^2$ = Convert N bits of Input 3 to polar form;
	\STATE b$_\mathrm{Nx1}^3$ = Convert N bits of Input 4 to polar form;
\end{algorithmic}

\begin{algorithmic}
\STATE {$------$ {Spread Process For Source 3} $------$}
\end{algorithmic}

\begin{algorithmic}[1]
\STATE A$_\mathrm{Nx1}^3$ = b$_\mathrm{Nx1}^2$.$*$c$_0$(:,1); $\rightarrow$ c$_0$ = [1 \quad 1]; 
\STATE A$_\mathrm{Nx1}^4$ = b$_\mathrm{Nx1}^2$.$*$c$_0$(:,2); $\rightarrow$ c$_0$ = [1 \quad 1];  	
\STATE S$_\mathrm{Nx2}^2$ = [A$_\mathrm{Nx1}^3$ \quad,\quad A$_\mathrm{Nx1}^4$]; 	
\end{algorithmic}

\begin{algorithmic}
\STATE {$------$ {Spread Process For Source 4} $------$}
\end{algorithmic}

\begin{algorithmic}[1]
\STATE B$_\mathrm{Nx1}^3$ = b$_\mathrm{Nx1}^3$.$*$c$_1$(:,1); $\rightarrow$ c$_1$ = [1 \hspace{1.6 mm} -1];
\STATE B$_\mathrm{Nx1}^4$ = b$_\mathrm{Nx1}^3$.$*$c$_1$(:,2); $\rightarrow$ c$_1$ = [1 \hspace{1.6 mm} -1];  	
\STATE S$_\mathrm{Nx2}^3$ = [B$_\mathrm{Nx1}^3$ \quad,\quad B$_\mathrm{Nx1}^4$]; 

\end{algorithmic}

\begin{algorithmic}
\STATE {$---------$ {Combining Process} $---------$}
\end{algorithmic}

\begin{algorithmic}[1]
\STATE P$_\mathrm{Nx2}^3$ = [A$_\mathrm{Nx1}^3$ \hspace{0.8 mm},\hspace{0.8 mm} B$_\mathrm{Nx1}^3$];
\STATE C$_\mathrm{Nx1}^3$ = $\frac{1}{2}$.$*$sum $\left(P_\mathrm{Nx2}^3 ,2 \right)$ ;$\rightarrow$ Combined step
\STATE P$_\mathrm{Nx2}^4$ = [A$_\mathrm{Nx1}^4$ \hspace{0.8 mm},\hspace{0.8 mm} B$_\mathrm{Nx1}^4$];
\STATE C$_\mathrm{Nx1}^4$ = $\frac{1}{2}$.$*$sum $\left(P_\mathrm{Nx2}^4 ,2 \right)$ ;$\rightarrow$ Combined step

\STATE x$_\mathrm{Nx2}^2$ =  [C$_\mathrm{Nx1}^3$ \hspace{0.8 mm},\hspace{0.8 mm} C$_\mathrm{Nx1}^4$];$\rightarrow$ The output of unit 2	
\end{algorithmic}
\begin{algorithmic}
	\STATE \textbf{Output of the M-STC Technique}
\end{algorithmic}
\begin{algorithmic}[1]
	\STATE D$_\mathrm{Nx1}^1$ = x$_\mathrm{Nx2}^1$(:,1) + j.$*$ x$_\mathrm{Nx2}^2$(:,1);
	\STATE D$_\mathrm{Nx1}^2$ = x$_\mathrm{Nx2}^1$(:,2) + j.$*$ x$_\mathrm{Nx2}^2$(:,2);
	\STATE X$_\mathrm{Nx2}^c$ =  $\left[\mathrm{D_{Nx1}^1} \quad, \quad \mathrm{D_{Nx1}^2} \right]$;
\end{algorithmic}		
\end{algorithm}

\begin{algorithm}[ht!]
	\caption{: The M-STE Technique}
	\label{M_STE_Tech}
	
	\begin{algorithmic}
		\STATE \textbf{Input of the M-STE Technique}
	\end{algorithmic}
	\begin{algorithmic}[1]
		\STATE {Y$_\mathrm{2Nx1}^c$ = Y$_\mathrm{2Nx1}^{real}$ + $j$Y$_\mathrm{2Nx1}^{imag}$;}	
	\end{algorithmic}	
	
	\begin{algorithmic}
		\STATE \textbf{First unit} $*******************************$
	\end{algorithmic}
	\begin{algorithmic}[1]
		\STATE Y$_\mathrm{2Nx1}^{real}$ =  real ( Y$_\mathrm{2Nx1}^c$) ;
		\STATE M$_\mathrm{Nx2}^{R}$ $\rightarrow$ Convert real part to a Nx2 Matrix form;
	\end{algorithmic}
	
	\begin{algorithmic}
		\STATE {$---------$ {Spread Process} $---------$}
	\end{algorithmic}
	
	\begin{algorithmic}[1]
		\STATE M$_\mathrm{Nx2}^{R0}$ = M$_\mathrm{Nx2}^{R}$.*c$_0$; $\rightarrow$ c$_0$ = c (1,:) = [1 \quad 1];
		\STATE M$_\mathrm{Nx2}^{R1}$ = M$_\mathrm{Nx2}^{R}$.*c$_1$; $\rightarrow$ c$_1$ = c (2,:) = [1 \hspace{1.6 mm} -1];
	\end{algorithmic}
	
	\begin{algorithmic}
		\STATE {$---------$ {Combining Process} $---------$}
	\end{algorithmic}
	
	\begin{algorithmic}[1]
		\FOR{$k1 = 1:N$}	
		
		\STATE D$_\mathrm{R}^0$(k1) = sum$\left( M_\mathrm{Nx2}^{R0}(k1,:)\right)$;		
		\STATE D$_\mathrm{R}^1$(k1) = sum$\left( M_\mathrm{Nx2}^{R1}(k1,:)\right)$;      
		\ENDFOR	
		\STATE D$_\mathrm{2Nx1}^R$ = $\left[D_\mathrm{R}^0 \quad , \quad D_\mathrm{R}^1 \right]$;
		\STATE D$_\mathrm{2Nx1}^R$ = \resizebox{0.2\columnwidth}{!}{$\frac{D_\mathrm{2Nx1}^R + 1}{2}$}
	\end{algorithmic}
	
    \begin{algorithmic}
     \STATE {$---------$ {Decision Block} $---------$}
   \end{algorithmic}
  \begin{algorithmic}[1]
    \FOR{$m1 = 1:2N$}	
    \IF {$D_\mathrm{2Nx1}^R(m1) > 0.5$}
    \STATE $D_\mathrm{2Nx1}^R(m1) = 1$
    \ELSE
    \STATE  $D_\mathrm{2Nx1}^R(m1) = 0$
    \ENDIF          
    \ENDFOR	
  \end{algorithmic}

	\begin{algorithmic}
		\STATE \textbf{Second unit} $*******************************$
	\end{algorithmic}
\begin{algorithmic}[1]
	\STATE Y$_\mathrm{2Nx1}^{imag}$ =  imag ( Y$_\mathrm{2Nx1}^c$) ;
	\STATE M$_\mathrm{Nx2}^{I}$ $\rightarrow$ Convert imaginary part to a Nx2 Matrix form;
\end{algorithmic}

\begin{algorithmic}
	\STATE {$---------$ {Spread Process} $---------$}
\end{algorithmic}
\begin{algorithmic}[1]
	\STATE M$_\mathrm{Nx2}^{I0}$ = M$_\mathrm{Nx2}^{I}$.*c$_0$; $\rightarrow$ c$_0$ = c (1,:) = [1 \quad 1];
	\STATE M$_\mathrm{Nx2}^{I1}$ = M$_\mathrm{Nx2}^{I}$.*c$_1$; $\rightarrow$ c$_1$ = c (2,:) = [1 \hspace{1.6 mm} -1];
\end{algorithmic}

\begin{algorithmic}
	\STATE {$---------$ {Combining Process} $---------$}
\end{algorithmic}

\begin{algorithmic}[1]
	\FOR{$k2 = 1:N$}	
	
	\STATE D$_\mathrm{I}^0$(k2) = sum$\left( M_\mathrm{Nx2}^{I0}(k2,:)\right)$;		
	\STATE D$_\mathrm{I}^1$(k2) = sum$\left( M_\mathrm{Nx2}^{I1}(k2,:)\right)$;      
	\ENDFOR	
	\STATE D$_\mathrm{2Nx1}^I$ = $\left[D_\mathrm{I}^0 \quad, \quad D_\mathrm{I}^1 \right]$;
	\STATE D$_\mathrm{2Nx1}^I$ = \resizebox{0.2\columnwidth}{!}{$\frac{D_\mathrm{2Nx1}^I + 1}{2}$}
\end{algorithmic}

\begin{algorithmic}
	\STATE {$---------$ {Decision Block} $---------$}
\end{algorithmic}
\begin{algorithmic}[1]
	\FOR{$m2 = 1:2N$}	
	\IF {$D_\mathrm{2Nx1}^I(m2) > 0.5$}
	\STATE $D_\mathrm{2Nx1}^I(m2) = 1$
	\ELSE
	\STATE  $D_\mathrm{2Nx1}^I(m2) = 0$
	\ENDIF          
	\ENDFOR	
\end{algorithmic}
\begin{algorithmic}
	\STATE \textbf{Output of the M-STE Technique}
\end{algorithmic}
\begin{algorithmic}[1]
	\STATE {D$_\mathrm{4Nx1}$ = $\left[D_\mathrm{2Nx1}^R \quad, \quad D_\mathrm{2Nx1}^I \right]$;}	
\end{algorithmic}	
	
\end{algorithm}
    \section{Conclusion}
The M-STC technique is proposed in this article as a promising technique. This technique decreases the OFDM symbol time to 25\% by saving 75\% of the OFDM symbol time. As a result, it reduces the  utilized bandwidth to a quarter, allowing the same amount of data to be transmitted using just a fourth of the bandwidth used by the OFDM system. This ultimately leads to increase the capacity. Additionally, it enhances system performance by minimizing the PAPR issue in OFDM systems. According to simulation results, utilizing C-STC-OFDM resulted in a 50\% reduction in processing time for an OFDM symbol when compared to a traditional OFDM system. While employing the proposed method M-STC-OFDM dramatically reduces the required time for each OFDM symbol by 75\% when compared to a traditional OFDM system.  In terms of PSD, the suggested system M-STC-OFDM decreases the BW by 75\% (BW$\mathrm{_{M-STC-OFDM}\sim }$ 45 kHz) when compared to the typical OFDM system (BW$\mathrm{_{OFDM}\sim}$ 180 kHz), while the C-STC-OFDM system in \cite{DoublingIoT} reduces the BW by 50\% (BW$\mathrm{_{M-STC-OFDM}\sim }$ 90 kHz). The PAPR is also reduced when the M-STC technique is used to an OFDM system. As the PAPR enhancement is 1.18 dB, or around 230\%, more than it was with the C-STC-OFDM system. The conventional OFDM system and the M-STC-OFDM system have the same BER (BER = $10^{-6}$) and SNR (SNR $\sim$ 10.5 dB) when compared to each other. Finally, The 16-QAM-OFDM system requires an SNR that is 3.8 dB higher than the M-STC-OFDM system in order to achieve the same BER (BER = $10^{-6}$).  As a result, the M-STC-OFDM system can transfer the same amount of data as the 16-QAM-OFDM system while consuming less power.

    \bibliography{main}

\end{document}